\documentclass[twocolumn,secnumarabic,amssymb, nobibnotes, aps, prl]{revtex4-2}

\usepackage{graphicx}
\usepackage[utf8]{inputenc}
\usepackage[english]{babel}
\usepackage{ textcomp }
\usepackage{natbib,hyperref}
\begin{document}

\title{Effect of electron correlations on attosecond photoionization delays \\ in the vicinity of the Cooper minima of argon}%

\author{D. Hammerland$^1$}
\thanks{equally contributed}
\author{P. Zhang$^1$}
\thanks{equally contributed}
\author{A. Bray$^2$}
\author{C. F. Perry$^1$} 
\author{S. K{\"u}hn$^3$}
\author{P. Jojart$^3$}
\author{I. Seres$^3$}
\author{V. Zuba$^3$}
\author{Z. Varallyay$^3$}
\author{K. Osvay$^3$}
\author{A. Kheifets$^2$}
\author{T. T. Luu$^1$} 
\email[Corresponding author: ]{trung.luu@phys.chem.ethz.ch}
\author{H. J.  W{\"o}rner$^1$}
\affiliation{$^1$Laboratorium f{\"u}r Physikalische Chemie, ETH Z{\"u}rich, Vladamir Prelog-Weg 2, Z{\"u}rich 8093, Switzerland.}
\affiliation{$^2$Research School of Physics, The Australian National University, Canberra ACT 0200,
Australia.}
\affiliation{$^3$Extreme Light Infrastructure Attosecond Light Pulse Source, Wolfgang Sandner utca 3, 6728 Szeged, Hungary.}
\date{June 2019}%

\begin{abstract}
Attosecond photoionization delays have mostly been interpreted within the single-particle approximation of multi-electron systems. The strong electron correlation between the photoionization channels associated with the 3p and 3s orbitals of argon presents an interesting arena where this single-particle approximation breaks down. Around photon energies of 42~eV, the 3s photoionization channel of argon experiences a ``Cooper-like" minimum, which is exclusively the result of inter-electronic correlations with the 3p shell. Photoionization delays around this ``Cooper-like" minimum have been predicted theoretically, but experimental verification has remained a challenge since the associated photoionization cross section is inherently very low. Here, we report the measurement of photoionization delays around the Cooper-like minimum that were acquired with the 100~kHz High-Repetition 1 laser system at the ELI-ALPS facility. We report relative photoionization delays reaching up to unprecedented values of 430$\pm$20~as, as a result of electron correlation. Our experimental results are in partial agreement with state-of-the-art theoretical results, but also demonstrate the need for additional theoretical developments. 
\end{abstract}

\maketitle
Cooper minima are the consequence of a sign change of the partial-wave photoionization matrix element of a given ionization channel as a result of energy-dependent radial overlap between the continuum and bound wave functions. For this reason, Cooper minima only arise when the radial wave function of the initial state has at least one node \cite{Cooper1962}. In the particular case of an s initial state, a single p partial-wave continuum can be accessed. This would normally lead to the cross section going through an exact zero at the photon energy at which the matrix element changes sign, and consequently a variation of its phase by $\pi$. All other initial states with angular momentum quantum number $\ell>0$ can be photoionized to two partial-wave continua ($\ell-1$ and $\ell+1$). The presence of a second ionization channel prevents the photoionization cross section from reaching zero and additionally reduces the phase shift to a value of less than $\pi$.

\begin{figure} 
\centering
    \includegraphics[width=.4\textwidth]{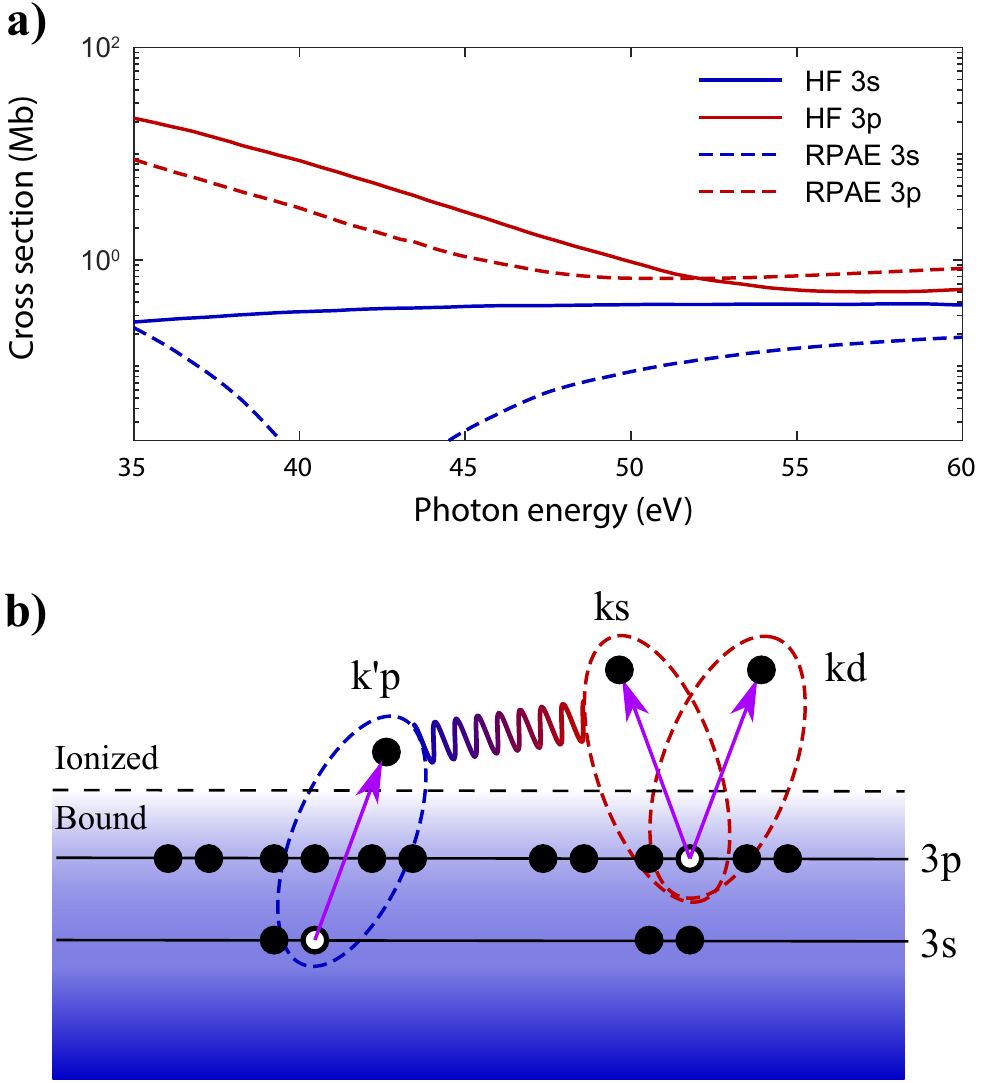}
\caption{a) Photoionization cross sections of Ar as calculated within the Hartree-Fock (solid) and the RPAE (dashed) approximations. Note that the HF calculations do not feature the 3s Cooper-like minimum at all. b) Illustration of the 3s-3p ionization channel coupling that causes the appearance of the Cooper-like minimum in the 3s photoionization channel.}
\label{ic}
\end{figure} 

The 3p ionization channel of argon experiences a well-known Cooper minimum around 50~eV, as shown in the red curves of Fig. \ref{ic}a). Within the single-particle approximation of multi-electron wave functions, such as the Hartree-Fock approximation, the 3s ionization channel features no Cooper minimum, as illustrated by the full blue line of Fig. \ref{ic}a). The inclusion of electron correlation, e.g. within the Random-Phase Approximation with Exchange (RPAE), leads to the appearance of a ``Cooper-like'' minimum around 42~eV in the 3s photoionization channel (dashed blue line in Fig. \ref{ic}a). The occurrence of this minimum is thus exclusively the consequence of electron correlation. In physical terms, the appearance of this minimum arises from dipole oscillations in the 3p ionization channel that effectively screen the 3s orbital from the incident light, thereby causing the appearance of the minimum \cite{Amusia1972}. The physical origin of the Cooper-like minimum is thus remarkably different from that of the usual Cooper minima. It is solely based on electron correlation and therefore represents one of the most remarkable manifestation of such effects.

\begin{figure} 
\centering
    \includegraphics[width=.5\textwidth]{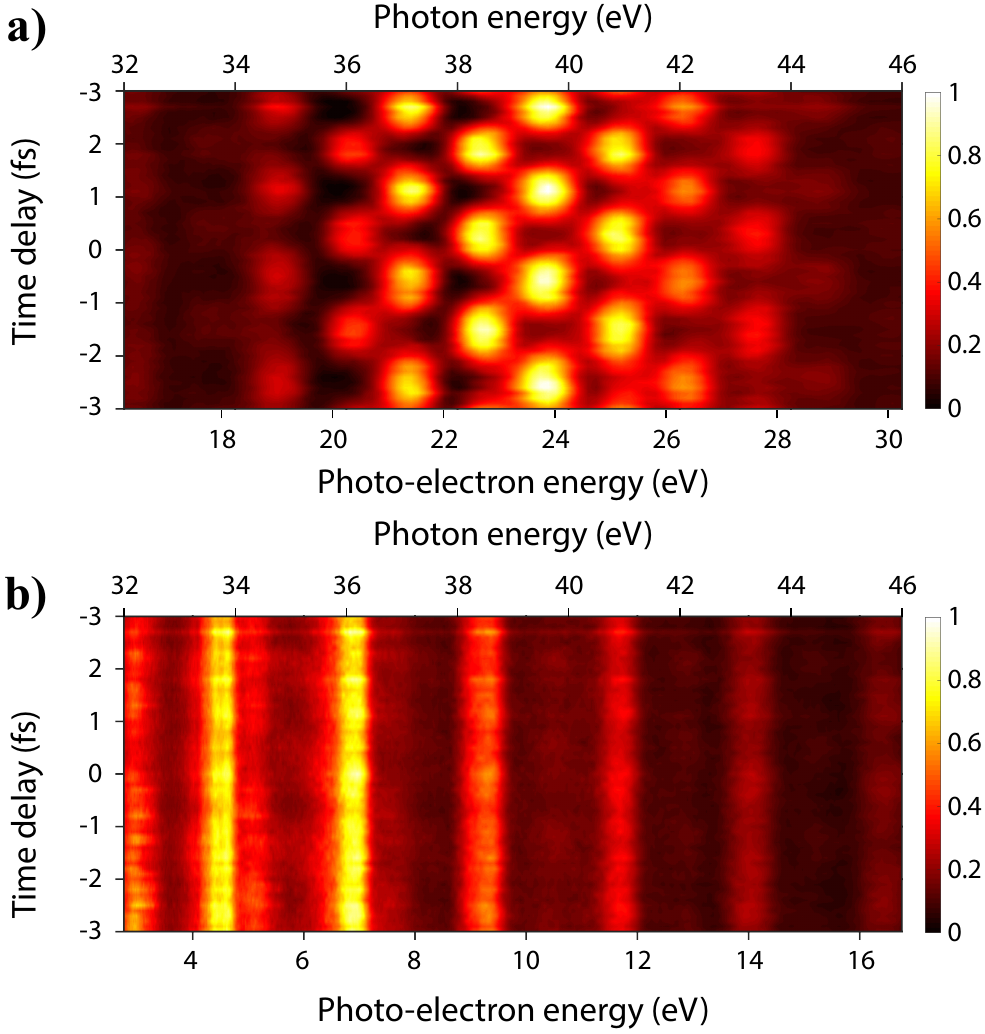}
\caption{Photoelectron interferogram from argon in the regions dominated by a) the 3p and b) the 3s photoelectrons. The horizontal axes show the photon energy (top) and the photoelectron kinetic energies of the dominant photoionization channel.} 
\label{ArPES}
\end{figure} 

Attosecond photoionization delays have proven to be a highly sensitive method for investigating electronic structure and dynamics on their fundamental time scales \cite{Schultze2010,Klunder2011,Pazourek2015a}. In the absence of strong channel interactions, these delays have been relatively well explained within single-particle approximations, such as Hartree-Fock calculations, e.g. in the case of most rare-gas atoms \cite{Kheifets2013,Guenot2014, Palatchi2014}. Additional effects, such as those arising from spin-orbit coupling \cite{Jordan2017} and shape resonances in molecules \cite{Huppert2016a,Baykusheva2017} could also be explained within the single-particle approximation. Photoionization delays in CO could also be approximately reproduced without invoking electron correlation \cite{Vos2018} and in the case of H$_2$, coupling between electronic and nuclear motion was found to be significant \cite{Cattaneo2018}. Certain specific observations of photoionization delays have however revealed the importance of electron correlation, e.g. delays in the vicinity of Fano resonances \cite{Gruson2016,Cirelli2018,Banerjee2019} or shake-up satellites \cite{Ossiander2017,Isinger2017}. The corresponding photoionization delays were however relatively small, typically ranging from a few to less than 50 attoseconds.

Since the first measurement of photoionization delays in argon \cite{Klunder2011}, its Cooper minima have attracted notable theoretical attention, although the original measurements did not cover the corresponding energy region. In particular, the RPAE \cite{Guenot2012}, Many Body Perturbation Theory (MBPT) method \cite{Dahlstrom2012a}, and the Relativistic Random Phase Approximation (RRPA) \cite{Saha2014a}, have all been applied to predict relative photoionization delays between the 3s and 3p channels. Qualitatively, these state-of-the-art methods all agree upon a large local increase of the 3p-3s relative photoionization delays from a few tens to several hundreds of attoseconds, i.e. the largest effect of electron correlation that has ever been predicted. Notably, these advanced theoretical techniques disagree, both with respect to the photon energies, and the absolute values of the predicted maximal delays. These differences reveal a remarkable sensitivity of the calculated delays to the detailed description of electron correlation in the different theoretical methods, additionally motivating the present work.

In this Letter, we report measurements of attosecond photoionization delays around the Cooper-like minimum in the argon 3s photoionization channel using the Reconstruction of Attosecond Beating By Interference of two-photon Transitions (RABBIT) technique \cite{Paul2001,Muller2002,Klunder2011}. The major experimental hurdle for measuring photoelectron spectra around this Cooper-like minimum is the exceptionally weak signal. This problem is exacerbated by the fact that most high-pulse-energy laser systems are bottlenecked to few-kHz repetition rates, leading to low acquisition rates. To overcome these limitations, our measurements utilized the 100~kHz High-Repetition 1 laser system at the Extreme Light Infrastructure Attosecond Light Pulse Source (ELI-ALPS) facility \cite{Kuhn2017}. The measurements were taken using $\sim$55~W of average power. The laser pulses were $\sim$40~fs in duration with a central wavelength of 1030~nm. The XUV mirror reflectivity was centered around 44~eV and had a bandwidth of $\sim$7~eV \cite{Hammerland2019}. A notable difference between our measurements and previous efforts \cite{Klunder2011, Guenot2012} is the utilization of a field-free time-of-flight spectrometer with an incorporated electromagnetic coil inside the time-of-flight tube to guide the electrons onto the multi-channel plate detector as opposed to a magnetic-bottle spectrometer. This enabled an increased collection efficiency without averaging over all emission angles, which has been predicted to greatly affect the measured delay \cite{Baykusheva2017,Bray2018}. For further details on the experimental setup, see \cite{Jordan2018,Hammerland2019}.

The attosecond photoelectron interferograms from the argon measurements were split into two separate interferograms for the 3p channel, Fig. \ref{ArPES}a), and the 3s channel, Fig. \ref{ArPES}b). The high count rate arising in the 3p channel relative to the 3s channel is a direct result of the Cooper-like minimum in the latter, which causes a two-order-of-magnitude decrease relative to the 3p channel, as can be seen in Fig.~\ref{ic}. The higher relative intensity of the side bands visible in panel a) of Fig. \ref{ArPES}, compared to panel b), is the result of the higher kinetic energy of the 3p photoelectrons \cite{Klunder2011,Dahlstrom2012a}. 

The photoelectron interferogram from Fig. \ref{ArPES}b) is influenced by a very weak ($\sim$1$\%$) local maximum in the XUV mirror reflectivity around 30 eV. As a consequence, the weak 3s-photoelectron signal generated by $\sim$43~eV photons are accompanied by 3p electrons ionized by $\sim$30~eV photons. However, the energy difference between the 3s and 3p ionization energies is not an integer multiple of the 1.2~eV dressing field, leading to a separation between the 3s and 3p electrons of $\sim$0.3~eV. This gap is well resolved by our photoelectron spectrometer at these low kinetic energies \cite{Jordan2018}, and is particularly visible at about 4~eV photoelectron energy in Fig. \ref{ArPES}b). This energy difference is most easily visible when a top-hat filter centered around the 2$\omega$ frequency is applied. Similar techniques have been used to study close-lying photoelectron bands in molecules \cite{Huppert2016a} and shake-up satellite peaks \cite{Isinger2017}.
The beating frequencies present in the side-band signals were extracted by Fourier transforming the photoelectron scan along the temporal-delay axis. The $2\omega$ oscillations were then selected by taking a line cut from the Fourier-transformed data. The sideband phases corresponding to the 3p and non-overlapping 3s photoelectrons were selected and subtracted. 

\begin{figure} 
\centering
    \includegraphics[width=.5\textwidth]{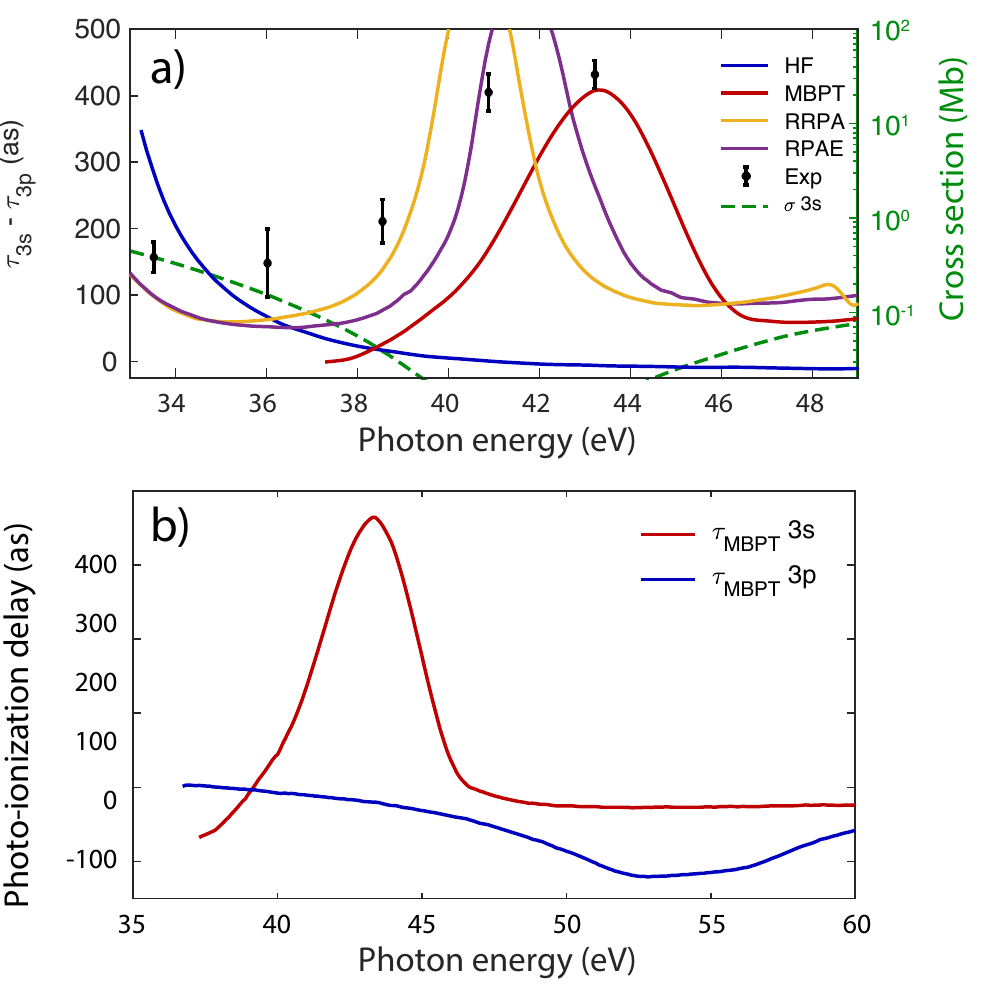}
\caption{a) Experimental data (black) compared to all published calculations (HF and MBPT from \cite{Dahlstrom2012a}, RPAE from \cite{Guenot2012}, RRPA from \cite{Saha2014a}), including the continuum-continuum delays, together with the 3s cross section (green). The error bars were determined by taking the square root of the deviations over the 3s and 3p errors added in quadrature. b) Atomic contributions ($\tau_{\rm W}$) . } 
\label{delay}
\end{figure} 

The measured relative photoionization delays \mbox{($\tau_{\rm{3s}}-\tau_{\rm{3p}}$)} are shown in Fig. \ref{delay}a), where they are compared to all previously published theoretical predictions (full lines) and the 3s photoionization cross section (dashed green line). The measured delays show a pronounced increase from $\sim$150~as at the lowest displayed photon energies to 430 $\pm$ 20~as in the vicinity of the Cooper-like minimum. Whereas this local maximum is entirely missing from the calculations neglecting electron correlation (HF method, blue line), all three calculations that explicitly include electron correlation (RPAE, RRPA and MBPT) predict local maxima in the correct energy range, yet none of them is in quantitative agreement with the experimental results. This qualitative agreement confirms electron correlation as the origin of the large observed relative delays. However, it also shows the necessity for further developments in the theoretical description of electron correlation and its implementation in calculations of atomic photoionization delays. Figure \ref{delay}b) shows the contributions of the atomic part (the Wigner delay, $\tau_{\rm W}$) of the photoionization delays, i.e. excluding the continuum-continuum contributions, to the measured difference $\tau_{\rm{3s}}-\tau_{\rm{3p}}$. For clarity, this comparison is restricted to the MBPT results. Whereas the Cooper-like minimum in the 3s channel leads to a local maximum at $\sim$43~eV, the Cooper minimum in the 3p channel results in a local minimum at $\sim$54~eV. The former thus represents the dominant contribution to our measured delays, whereas the latter is less important.

The demonstrated ability to measure and explain correlation-induced delays of several hundreds of attoseconds in the vicinity of anti-resonances (the Cooper and Cooper-like minima) provides the valuable opportunity to question the separability of attosecond photoionization delays $\tau$ into an atomic contribution (the Wigner delay) $\tau_{\rm W}$ and a universal continuum-continuum delay $\tau_{\rm cc}$. This separability is well established for atoms in the absence of resonances and results from the independence of $\tau_{\rm cc}$ on the angular-momentum quantum number $\ell$ of the involved partial-wave continua \cite{Dahlstrom2012a}. In the case of molecules, this separability breaks down because of quantum interference between two-photon two-color photoionization pathways involving different values of $\ell$ \cite{Baykusheva2017}. The separability may therefore also be expected to break down in atoms in the vicinity of (anti-)resonances where the relative phase of partial-wave matrix elements undergo rapid changes. Moreover, the possible failure of this separability in the presence of channel interactions (i.e. electron correlations) has also not been verified previously.

In the present work, we investigate this question by comparing the RPAE results for the delays in the Ar 3p channel with the explicit solution of the one-electron time-dependent Schr\"odinger equation (TDSE) in three dimensions in the presence of the ionizing XUV and dressing IR fields. The TDSE is solved in the presence of an effective potential that has been shown to accurately reproduce both the ground state energy and the location of the Cooper minimum \cite{Woerner2009}. The IR and XUV fields were taken from the experimental parameters and the photoionization delays $\tau_{\rm 3p}$ (Fig. \ref{delay}) were extracted using infinite-time surface flux methods \cite{Morales2016}. For a more detailed description of the TDSE methods used in this work, the reader is referred to \cite{Bray2018}. 

\begin{figure} 
\centering
    \includegraphics[width=.5\textwidth]{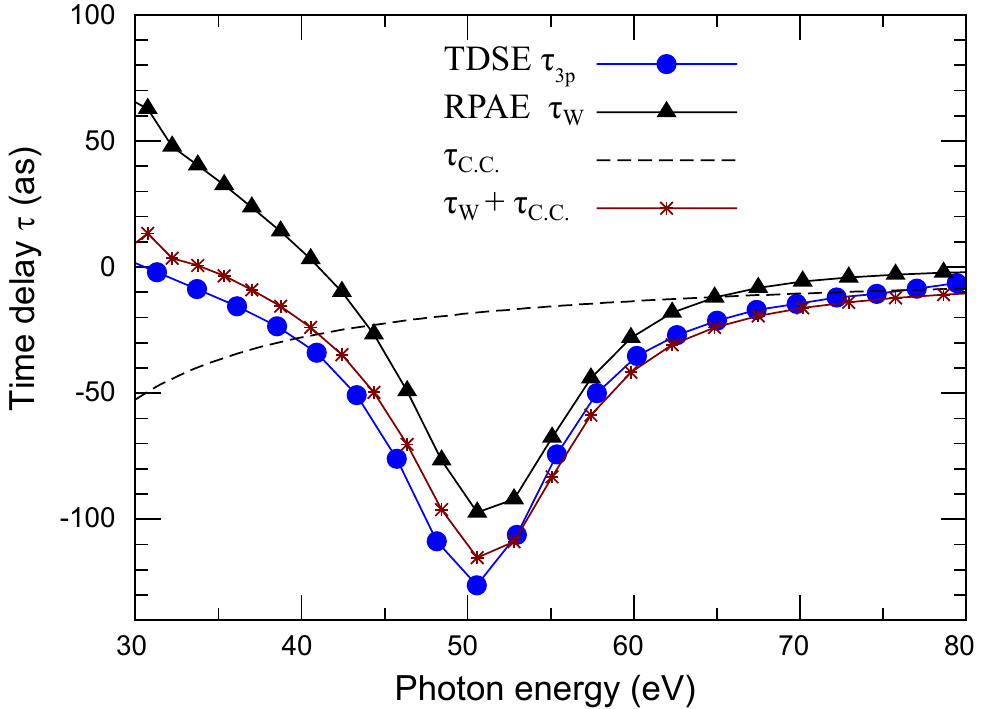}
\caption{3p photoionization delays of argon calculated with the TDSE (blue) or RPAE methods (black). With the inclusion of the continuum-continuum delays into the RPAE calculations (red), the two results agree within the computational error of 10 as.} 
\label{TDSE}
\end{figure}

Figure \ref{TDSE} compares the results of the TDSE calculations with those of the RPAE calculations. The atomic delay is shown as black triangles and the analytical continuum-continuum delay is shown as a dashed line. The sum of the two ($\tau_{\rm W}+\tau_{\rm cc}$) is shown as brown stars. Above the location of the Cooper minimum, the two theories are in excellent agreement. Below the Cooper minimum, the deviation amounts to less than 10 attoseconds, which corresponds to the expected accuracy of the two types of calculations. This level of agreement shows that the separability of atomic delays into Wigner and continuum-continuum delays still holds, at least in the case of the argon Cooper minimum, excluding the possibility of a systematic breakdown of the separability in the vicinity of atomic resonances. The remaining discrepancies may result from the single-active-electron approximation used in the TDSE calculations and/or the approximations made within the RPAE calculations, such as the neglect of doubly-excited configurations, which has previously been mentioned as a possible source of deviations between experiment and theory \cite{Jordan2017}. 

In conclusion, we have measured photoionization delays around the Cooper-like minimum of argon. Our measurements have revealed the largest relative photoionization delays measured to date, reaching up to 430 $\pm$ 20~as. Comparison with state-of-the-art calculations has revealed that electron correlation is the main source of these remarkably large delays. Our experimental results have validated these advanced calculations and confirmed the dramatic impact of channel interactions on photoionization delays. The mutual agreement of RPAE and TDSE calculations on photoionization delays in the vicinity of the Cooper minimum has validated both approaches and confirmed the separability of atomic photoionization delays into a Wigner-like delay and a continuum-continuum delay, even in the vicinity of \mbox{(anti-)~resonances}. These results establish the measurement of attosecond photoionization delays as a very sensitive approach to revealing the role of electron correlations and a highly sensitive technique for testing the most advanced electronic-structure methods.

The authors acknowledge all of the workers at the ELI-ALPS laser facility for their dedicated work throughout the beam times, particularly Dr. Tam{\'a}s Csizmadia and Dr. Mikl{\'o}s F{\"u}le, the preparatory efforts of Dr. Arohi Jain and Dr. Thomas Gaumnitz for the realization of this work and the technical staff of the ETH Z{\"u}rich Laboratorium f{\"u}r Physikalische Chemie, especially Andreas Schneider for help with electronics as well as Mario Seiler and Markus Kerellaj for the custom-manufactured mechanical parts used during the experiment. Finally, the authors also acknowledge Dr. Adam Smith and Danylo Matselyukh for invaluable discussions.

This research was funded by ETH Z{\"u}rich and the Swiss National Science Foundation through grants 200021E\_162822 and 200021\_172946. Resources of the National Computational Infrastructure were employed.  

\bibliography{ArCooperMinV8.bib}

\end{document}